\documentclass[usenatbib]{basi}
\usepackage[british]{babel}
\usepackage[varg]{txfonts}
\usepackage{rotating}
\usepackage{dcolumn}
\begin{document}
\title[Low-frequency GMRT observations of the magnetic Bp star HR Lup]
{Low-frequency GMRT observations of the magnetic Bp star HR Lup (HD\,133880)}
\author[S.J. George and I.R. Stevens]%
       {Samuel~J.~George$^{1,2}$\thanks{email: \texttt{sgeorge@mrao.cam.ac.uk}} and Ian~R.~Stevens$^{2}$\\
       $^1$Astrophysics Group, The Cavendish Laboratory, JJ Thomson Avenue, University of Cambridge, \\ 
           Cambridge, CB3 0HE, UK\\
       $^2$Astrophysics and Space Research Group, School of Physics and Astronomy, University of Birmingham, \\ 
           Birmingham, B15 2TT, UK \\
}

\pubyear{2012}
\volume{40}
%\pagerange{\pageref{firstpage}--\pageref{lastpage}}
\status{in press}

\date{Received 2012 May 03; accepted 2012 May 24}

\maketitle
%------------------------------------------------------------------------------%
% abstract and keywords                                                        %
%------------------------------------------------------------------------------%
\label{firstpage}
%------------------------------------------------------------------------------%
\begin{abstract}
We present radio observations of the {magnetic} chemically peculiar Bp star HR Lup (HD 133880) 
at $647$ and {$277$~MHz} with the GMRT. At both frequencies the source is not detected but we 
are able to determine upper limits to the emission. The {$647$\,MHz} limits are particularly 
useful, with a  $5\sigma$ value of 0.45\,mJy.  Also, no large enhancements of the 
emission were seen. The non-detections, along with previously published higher frequency detections, 
provide evidence that an optically thick gyrosynchrotron model is the correct mechanism for 
the radio emission of HR Lup. 
\end{abstract}
%------------------------------------------------------------------------------%

%------------------------------------------------------------------------------%
% main text of the paper, using \section, \subsection, \subsubsection          %
%------------------------------------------------------------------------------%
\section{Introduction}\label{s:intro}

Sources of stellar radio emission can be split by the presence of
non-thermal emission. Thermal emission is seen in massive stars
(Wolf-Rayet, O- and early B-type stars) that have strong radiatively
driven stellar winds \cite[]{Wright1975}. These winds are ionized and
radiate via free-free emission. A number of early type stars also
show a non-thermal component and this is usually due to
shock-acceleration associated with colliding stellar winds in binary
systems \cite[]{debecker2007}.  Several types of lower mass stars also
show detectable radio emission, often non-thermal in nature. Examples
include the Sun, RS CVn stars \cite[]{slee2008}, dMe stars
\cite[]{Osten2006}, {dwarf stars and brown dwarfs} \cite[]{Berger2006}. This emission
is usually associated with coronal emission and flare activity, is
often time variable and also requires the presence of magnetic fields.

Late B-type stars and early A-type stars have neither strong stellar
winds nor substantial outer convective zones and so were not
expected to be strong radio emitters. Magnetic chemically peculiar
(MCP) stars have been known as radio emitters in the centimetre range
since the mid 1980s. MCP stars are a class of peculiar A and B stars
(referred to as Ap or Bp stars) with strong (kGauss) magnetic fields.
The class is characterized by large abundance anomalies in a range of
elements which suggested that the photospheres of these stars are
highly stable (with the turbulent motions stabilized by the magnetic
field) allowing diffusion to take place. The origin of the magnetic
fields in these stars is still unclear, with possibilities including either 
a fossil-field from the original collapse to form the star, or possible
dynamo action either in the radiative zone or at an earlier stage of
evolution \cite[]{Arlt2008}. Although sub-surface convection zones in
massive stars, associated with the iron opacity bump, could be
associated with magnetic field generation, this mechanism 
{is thought to not work for the late B-type stars considered here} 
\cite[]{Cantiello2011}.

There is one MCP star that is of particular interest at radio
wavelengths -- CU Vir (A0p). This nearby star has a period of 0.52
days and a strong magnetic field. From optical variability timing,
\cite{Pyper1998} reported on a peculiar change in the optical (and
presumably rotational) period. CU Vir also shows periodic, polarised
outbursts as well as quiescent emission. The burst emission has been
attributed to electron cyclotron maser emission. The outbursts are
seen over a wide range of frequencies (see \cite[]{Lo2012};
\cite[]{Trigilio2011}; \cite[]{Ravi2010}; \cite[]{George2010} and references
therein).

{There have been a small number of radio observations of late B and
early A-stars. \cite{Drake1987} observed 34 sources detecting only 5
at $5$~GHz with the Very Large Array (VLA). They interpreted the
observed radio emission as being due to gyrosynchrotron emission from
the magnetosphere of the star due to a continuously injected
population of mildly relativistic particles that are trapped in the
magnetosphere. \cite{Wilson1988} also using the VLA at $5$~GHz surveyed  
16 M dwarf stars detecting one source, Gliese 735.
A more recent VLA survey of MCP stars at $5$~GHz by
\cite{Leone1994} extended this to cover 40 stars, of which only 8 were
detected.}

MCP stars on average show moderate circular {polarisation}, with radio 
properties similar to that of active cool stars such as RS CVns. 
They show periodic changes in the detected magnetic field over the 
stellar rotation period {\cite[]{Borra1980}}. The magnetic field topology 
of this kind of star is generally taken as a magnetic dipole tilted with 
respect to the rotation axis \cite[]{Babcock1949}, though in the case 
of HR Lup the field is quadrupolar in nature (see below). In a few 
cases, MCP stars possess an anisotropic stellar wind as a consequence 
of the wind interaction with the dipolar magnetic field \cite[]{Shore1990}. 
In about 25\% of MCP stars non-thermal radio emission is observed and 
the rate of detection seems to be correlated with the effective stellar 
temperature \cite[]{Linsky1992,Leone1994}. 

The radio emission is generally 
interpreted as gyrosynchrotron emission from mildly relativistic electrons. 
These electrons are accelerated in current sheets, formed when the gas flow 
brakes the magnetic field lines, close to the magnetic equator and the 
electrons propagate along the magnetic field lines towards the inner 
magnetospheric regions \cite[]{Havnes1984}. 

X-ray studies have found that MCP stars are weak X-ray emitters with a 
detection rate of just over 10\% \cite[]{drake1994}. The X-ray emission 
of MCP stars does not correlate with other stellar properties. Their 
radio properties also do not follow the Guedel-Benz scaling law as this 
would imply X-ray luminosities as high as $10^{33}$~erg~s$^{-1}$ 
\cite[]{guedel1993}. The only X-ray detections of MCP stars are not 
particularly bright \cite[$L_{x} > 10^{30.5}$~erg~s$^{-1}$; ][]{drake2006}. 

In this paper we focus on low frequency observations of one specific 
object (HR Lup) which may give some insights into the processes going 
on in the star.

%------------------------------------------------------------------------------%
\section{HR Lup}\label{s:hrlup}

HR Lup (HD~133880, HR~5624; RA: 15 08 12.124, Dec: $-$40 35 02.15) is a
rapidly rotating B-type chemically peculiar star of spectral type B8\,Ivp. 

The star is rapidly rotating ($v \sin{i} \approx 103$~kms$^{-1}$) and
has a rotational period of 0.8777~days, which is observed in optical
photometry, magnetic field measurements and radio flux
\cite[]{Bailey2012,schmitt2005}.

The inferred stellar parameters for HR Lup are $M_\ast=3.2M_\odot$,
$R_\ast=2.01 R_\odot$, $T_{eff}=13000$~K and age of 15.8\,Myr 
\cite[derived from the fact that it is a member of the Upper Cen Lup association; ][]{Landstreet2007}.
 HR~Lup has a very strong magnetic field,
typically $2.4$~kG \cite[]{schmitt2005}.  This magnetic field varies
from about 4 to +2 kG \cite[]{Landstreet1990}. Unlike most MCP stars
the magnetic field is not dipolar, but quadrupolar (and this has
consequences for the observed radio variability). The rotational and
magnetic axes are misaligned \cite[]{Bailey2012}.

HR Lup is a photometric variable with variations on the order of
$0.15$~mag in the U -band, which is probably the result of the large
magnetic field and surface abundance anomalies \cite[]{Waelkens1985}.

HR Lup is a known radio source, previously observed with the 
{Australia Telescope Compact Array (ATCA)}. 
At $5$~GHz \cite{lim1996} 
demonstrated that both the total intensity and
circular {polarisation} of the source varied significantly and
coherently according to the known rotational period. 
{The total intensity varied from $\sim1$~mJy to $\sim5$~mJy 
with the degree of circular {polarisation} reaching up to $20\%$.}
{They noted that the emission shows broad peaks (suggesting a dipole 
contribution to the field) and narrower peaks at the predicted
phases of a quadrupole contribution to the magnetic field.}
At $8$~GHz the source is seen to have a flux density of
$4.08 \pm 0.16$ mJy {though no indication of any variability 
is presented} \cite[]{drake2006}. {\cite{Bailey2012} 
have re-reduced the ATCA radio data of \cite{lim1996} but there 
are no significant differences in the reduced data.}

In this paper we present observations at {$647$ and $277$~MHz} with the intention of 
finding if there is any lower frequency emission and help to find if 
there is a cut-off frequency.

%------------------------------------------------------------------------------%
\section{Observations and data reduction}\label{s:gmrt}

The Giant Metrewave Radio Telescope (GMRT) is located near Pune, India and consists 
of thirty $45$~m diameter radio dishes. The GMRT has a maximum and minimum baseline 
of $25$~km and $100$~m, and has operating receivers at $150$, $235$, $325$, $610$ 
and $1400$~MHz. 

We observed HR Lup with the GMRT on 2009 December 5th and 7th simultaneously at 
both {$647$} and {$277$~MHz}, implying only total intensity maps {were} constructed. 
At 647~MHz a bandwith of 16~MHz was used and at 277~MHz a bandwith of 6~MHz was used 
(both observations used 128 channels).
{An integration time of 16 seconds was used}. The 
total time on source was $4.67$ hours over the two nights. {During the observations 27 
antennas were used}. The observations consisted of flux density and bandpass calibration 
using 3C\,286 
at the start and end of the observations, and  phase calibration using {VSOP J1501--3918} 
every $30$ minutes. 

Each spectral window was calibrated and imaged separately using the Common Astronomy Software 
Applications package (CASA\footnote{http://casa.nrao.edu/}).
{Any interference in the data was removed by manual inspection}. {Several rounds of phase 
only self-calibration were completed on the target data.
During these iterations any visibility measurements that showed unusual phase excursions 
were rejected.} The final image used for analysis below was primary beam corrected and has 
a central rms of $90~\mu$Jy at {$647$}~MHz and $6$~mJy at {$277$~MHz}. 

The source positions for the brighter sources were matched with that
of the Sydney University Molonglo Sky Survey (SUMSS) at $843$~MHz
\cite[]{Mauch2003}. No significant source position offset was found
between the GMRT data and this survey.
	
\section{Results and  Discussion}\label{s:results}

At {$647$}~MHz we do not detect the source. At the $5\sigma$ level this
gives us an upper limit of the flux density of $0.45$~mJy. Also, at {$277$~MHz}
we do not detect the source. At the $5\sigma$ level this gives us an
upper limit of the flux of $30$~mJy. 
{The data were also cut into shorter time-scales (at $120$ seconds and the 
scan length $\sim600$s) and imaged. No significant time variation was seen at
the position of the source thus ruling out any bright bursts.}

\cite{Bailey2012} have produced the most recent and 
accurate ephemeris for HR Lup, combining several data sets to yield:

\begin{equation}
{\rm JD}_{\rm min}= 2445472.000(10)+0.877476(9)\cdot E,
\label{ephemeris} 
\end{equation}
\noindent where zero phase corresponds to minimum photometric brightness and
also the minimum of the longitudinal magnetic field. With this
ephemeris the principal radio maxima at GHz frequencies occur at phase
$\phi = 0.0$ and $0.5$.

From this, we can determine the phase of our observations. The
observations are on two separate days and they cover phases of
$0.93$--$1.06$ and $0.15$--$0.35$ respectively. Thus the observations cover at
least one of the periods of radio maxima previously observed. Using
the $3\sigma$ period errors quoted by \cite{Bailey2012} we estimate
an uncertainty in the phase for these observations of around 0.1.

Since we compared our positions to that of SUMSS it is worth noting
that no source is given in their catalogue at the position of HR Lup, though
investigation of the SUMSS mosaics indicates that there is a possible
enhancement with a peak of $4.5$~mJy/beam at the optical position of
HR Lup. This source is below the detection threshold of $10$~mJy/beam.

If indeed there is a source corresponding to $4.5$~mJy/beam at
843\,MHz then this would imply an extremely steep cut-off between the
$843$ and {$647$} MHz observations (i.e. $\alpha\geq 7$, for
$S_\nu\propto \nu^{\alpha}$, which seems rather extreme). Though with
no formal detection in the SUMMS maps this does not discount the
possibility that the cut-off frequency is not located somewhere
between {$647$\,MHz} and the previous detections at $5$~GHz. If we
discount the SUMSS point, then the implied limit on the spectral index
between {$647$\,MHz} and the previously reported 5\,GHz points is
$\alpha>1$. These sources are possibly variable in time (although the
ATCA observations show only a factor 2--3 variability across the
rotational period) {over longer time-scales}.

{The non-detection, particularly at {$647$\,MHz}, implies 
that the radio emission region intersecting the line of sight 
has a finite extent which is consistent with the 
optically thick gyrosynchrotron model of \cite{Linsky1992}.}
The radio spectrum from this model will have a frequency ($\nu_{peak}$) where
the emission peaks, the location of which depends on the wind
density/geometry and the magnetic field properties. The radio
spectrum then falls away on either side of this peak frequency, and
the spectral slope on either side of the $\nu_{peak}$ with the
electron energy spectrum. It is likely that $\nu_{peak}$ lies
somewhere between {$647$\,MHz} and $5$~GHz, and observations with a much
broader bandwidth will be important to fully constrain the detailed
shape of the emission.

%------------------------------------------------------------------------------%
\section{Summary}\label{s:conclusions}

In summary, we present non-detections of the MCP star, HR Lup with the
GMRT at {$647$\,MHz} and {$277$~MHz}. No significant time variation was seen at
the position of the source thus ruling out bright bursts, as seen in
CU Vir, in this dataset. We consider that these non-detections provide
evidence that the emission mechanism for this star is optically thick
gyrosynchrotron emission. We suspect that the peak emission frequency
lies between {$647$\,MHz} and $5$~GHz. Follow-up observations at frequencies
in this range would be particularly useful to constrain the spectral
model and indeed provide constraints on the energy injection of
electrons in the stellar magnetosphere. Broad-band observations
(covering both the GHz and sub-GHz regimes) are necessary to constrain
the emission mechanisms in MCP stars and coverage of the entire
rotational period are necessary to detect the presence of short-lived
intense bursts, which may well be a feature of the radio emission from
the strongly magnetic stars.

%------------------------------------------------------------------------------%
\section*{Acknowledgements}

We thank the GMRT staff who have made these observations possible. 
The GMRT is run by the National Centre for Radio Astrophysics of the 
Tata Institute of Fundamental Research. 
{We want to thank the anonymous reviewers for their constructive reviews. 
The comments were a great help to improve the manuscript}.

\begin{table}
\begin{center}
\caption[Observations]{The GMRT observations of HR Lup in UT at {$647$/$277$~MHz}. 
Given are the start and end times of the observing and the range of phase covered by 
these observations, using the ephemeris of \cite{Bailey2012}.} \smallskip
\begin{tabular}{lcc}
\hline\\[-3mm] % newline after \hline, but with -ve space
Start Observing & End Observing & Phase range \\
\hline\\[-3mm] % newline after \hline, but with -ve space
04:35:10.9 05 Dec 2009 & 07:22:40.4 05 Dec 2009 & $0.93$--$1.06$\\
03:23:28.3 07 Dec 2009 & 07:35:41.3 07 Dec 2009 & $0.15$--$0.35$\\
\hline\\[-3mm] % newline after \hline, but with -ve space
\end{tabular}
\label{uss}
\end{center}
\end{table}

%------------------------------------------------------------------------------%
% bibliography: produced from ADS using custom format of                       %
%     %z132 \\bibitem[%\2%(y)%\3m]%{R}\n   %\8.1g,%\Y,%\q,%\V,%\ p             %
%------------------------------------------------------------------------------%

\label{lastpage}
%------------------------------------------------------------------------------%
\end{document}